\documentclass[12pt,preprint]{aastex}

\def\gtorder{\mathrel{\raise.3ex\hbox{$>$}\mkern-14mu
             \lower0.6ex\hbox{$\sim$}}}

\def\ltsima{$\; \buildrel < \over \sim \;$}
\def\simlt{\lower.5ex\hbox{\ltsima}}
\def\gtsima{$\; \buildrel > \over \sim \;$}
\def\simgt{\lower.5ex\hbox{\gtsima}}

\begin{document}


\title{The Progenitors of Short-Hard Gamma-Ray Bursts from an
Extended Sample of Events}


\author{Avishay Gal-Yam\altaffilmark{1}, Ehud Nakar, Eran O. Ofek,
S. B. Cenko, S.~R.~Kulkarni, A.~M.~Soderberg, F.~Harrison}
\affil{Division of Physics, Mathematics and Astronomy, California Institute of
     Technology, Pasadena, CA\,91125}
\author{D.~B.~Fox}
\affil{Department of Astronomy \& Astrophysics, 525 Davey Laboratory,
The Pennsylvania State University, University Park, PA 16802}
\author{P.~A.~Price}
\affil{Institute for Astronomy, University of Hawaii, 2680 Woodlawn Drive, Honolulu, HI\,96822}
\author{B.~E.~Penprase}
\affil{Department of Physics and Astronomy, Pomona College,
Claremont, CA\,91711}
\author{Dale A. Frail}
\affil{National Radio Astronomy Observatory, P.O. Box O, Socorro, NM 87801}
\author{J. L. Atteia}
\affil{Laboratoire d'Astrophysique de Toulouse-Tarbes, OMP, 31400 Toulouse, France}

\and

\author{E. Berger\altaffilmark{1}, M. Gladders, J. Mulchaey}
\affil{Observatories of the Carnegie Institution of Washington,
813 Santa Barbara Street, Pasadena, CA 91101}

\email{avishay@astro.caltech.edu, udini@tapir.caltech.edu}


\altaffiltext{1}{Hubble Fellow.}


\begin{abstract}

The detection and characterization of the afterglow emission and
host galaxies of short-hard gamma-ray bursts (SHBs) is one of the
most exciting recent astronomical discoveries. In particular,
indications that SHB progenitors belong to old stellar populations,
in contrast to those of the long-soft GRBs, provide a strong clue
about the physical nature of these systems. Definitive conclusions
are currently limited by the small number of SHBs with known hosts
available for study. Here, we present our investigation of SHBs
previously localized by the interplanetary network (IPN) using new
and archival optical and X-ray observations. We show that we can
likely identify the host galaxies/clusters for additional two
bursts, significantly expanding the sample of SHBs with known hosts and/or
distances. In particular, we determine at a very high probability $>3\sigma$
that the bright SHB 790613 occurred within the rich galaxy cluster
Abell 1892, making it probably the nearest SHB currently known. We
show that the brightest galaxy within the error box of SHB 000607,
at $z=0.1405$, is most likely the host galaxy of this event. 
Additionally, we rule out the existence of
galaxy overdensities (down to $\approx21$ mag) near the locations of
two other SHBs, and set a lower limit on their probable redshift. We
combine our SHB sample with events discovered recently by the {\it
Swift} and {\it HETE-2} missions, and investigate the properties of
the extended sample. This sample enables us to determine that the
progenitors of SHBs are typically older than these of type Ia SNe,
implying a typical life time of several Gyr. We also show that it is
unlikely that there is a significant population of progenitors with
life time $\lesssim 1$Gyr. This result is difficult to reconcile 
with the popular model of neutron-star mergers as the progenitors of SHBs.
We note that long SHB life times, if confirmed, imply that very few
such events occur above $z=1$, and that those above $z=0.5$ should preferably
occur in galaxy clusters. The low typical redshift of SHBs leads to 
a significant increase in the local SHB rate, and bodes well for the
detection of gravitational radiation from these events with forthcoming 
facilities, should they result from compact binary mergers. 
  
\end{abstract}




\section{Introduction}
The short-hard class of the Gamma-Ray Burst (GRB) population
(\citealt{Kouveliotou93}; SHB hereafter) makes up a quarter of the
entire GRB population observed by
BATSE\footnote{http://www.batse.msfc.nasa.gov/batse/} with an
all-sky detection rate of $\approx 170 \;\rm y^{-1}$
\citep{Meegan97}. For many years, the failure to detect any
afterglow emission associated with SHBs prevented rapid progress in
this field. Still, population analysis of  SHBs provided indirect
clues about their typical distance. Namely, the almost-isotropic sky
distribution \citep{Briggs96,Balazs98,Magliocchetti03} and small
value of $<V/V_{max}>$  \citep{Katz96,Schmidt01,Guetta05} suggested
a cosmological origin.

In the last few months the long-expected breakthrough in this field
has finally occurred, facilitated by accurate localizations of SHBs
by the {\it
Swift}\footnote{http://swift.gsfc.nasa.gov/docs/swift/swiftsc.html}
and {\it HETE-2}\footnote{http://space.mit.edu/HETE/Welcome.html}
spacecrafts. The discovery of the X-ray afterglow of GRB 050509b by
the {\it Swift} X-ray Telescope (XRT; \citealt{Gehrels05}) led to
its localization to within a few arc-seconds, in close proximity to
a bright elliptical galaxy, a member of a galaxy cluster at $z=0.22$
\citep{Bloom05,Kulkarni05,Castro05a,Gehrels05,prochaska05}. Sensitive follow-up
imaging of the XRT localization revealed many background galaxies
within the error circle but no optical afterglow \citep{Kulkarni05,
Gehrels05, Bloom05, Hjorth05a,Castro05a} and the lack of
sub-arcsecond localization prohibits an unambiguous identification
of the host galaxy of this SHB. However, a posteriori statistical
arguments suggest that SHB 050509b was associated with the $z=0.22$
system \citep{Gehrels05, Bloom05, Eisenstein05}.

The accurate localization of SHB 050709 by {\it HETE-2}
\citep{Butler05} and the subsequent discovery of X-ray
\citep{Fox05} and optical \citep{Price05,hjorth05} afterglow emission
pinpointed the location of this SHB to sub-arcsecond
accuracy, and led to the
identification and detailed characterization of its host galaxy
\citep{Fox05,Covino05,prochaska05}. This identification established that SHB
050709 is a relatively nearby event (at $z=0.16$), and that its host
is not an early-type galaxy, as suggested for 050509b, but rather
a modestly star-forming galaxy, with similar star formation rate
to Sb/c galaxies.

Shortly after, SHB 050724 was localized by {\it Swift}, and turned
out to have a rich afterglow spectrum including X-ray
(\citealt{Romano05,Fox05}), radio \citep{Cameron05,Berger05} and
optical/IR \citep{Gal-Yam05, Berger05, Castro05b,Cobb05,Wiersema05}.
The afterglow detection led to an unambiguous association of this
burst with a red early-type galaxy \citep{Berger05,prochaska05} at
$z=0.257$. A few weeks later, SHB 050813 was localized by {\it
Swift} and associated with a galaxy cluster at $z=0.722$
(\citealt{Gladders05,berger05b,prochaska05}; Berger et al., in preparation). These two
events give further credence to the association of SHB 050509b with the
nearby $z=0.22$ cluster elliptical.

These recent observations suggest, when taken together, that a
significant fraction, perhaps even all of the progenitors of SHBs
are drawn from old stellar populations, and are therefore
long-lived. This result establishes the physical difference
between SHB progenitors and the short-lived, massive star
progenitors of long-soft GRBs.

A few months before these exciting discoveries were made, the study
of SHBs was briefly revitalized by a Galactic event - the unusually
bright super-giant flare from a Soft Gamma-ray Repeater (SGR
1806-20). The temporal structure and luminosity of this event
indicated it would have appeared as an SHB had it occurred in a
nearby galaxy \citep{Duncan01,Dar05,Nakar05,Palmer05,Hurley05}, and
this possibility raised a flurry of speculations about the possible
association of SHBs with extra-galactic SGRs. In \citet{Nakar05} we
investigated this question by looking for bright host galaxies in
the error boxes of well-localized SHBs, expected if these were
indeed SGRs in nearby galaxies. This observational test has not been
attempted previously. This work set lower limits on the distance and
energy output of six SHBs and showed that only a small fraction of
SHBs can in fact be extra-galactic SGRs. The latter result was
independently confirmed, using other approaches, by
\citet{Palmer05,Popov05} and \citet{Lazzati05}. Recent analysis by
\citet{tanvir05} reveals an apparently significant correlation between
SHBs from the BATSE catalog and nearby galaxies, with the correlation 
getting stronger when galaxies of earlier type are considered. Since known
SGRs are observed in star-forming regions and are generally assumed to 
be young neutron stars, this result further indicates that SGRs do not
contribute significantly to the SHB population, even in the nearby 
Universe. These findings are
consistent with results from the analysis of BATSE SHBs mentioned
above, indicating a cosmological origin of SHBs.

Our mostly archival
investigation has shown the wealth of information that can be
extracted from arcmin$^2$ error boxes of SHBs, especially when deep
observations of the error boxes are obtained. Detailed exploration
of the error boxes, especially in view of the breakthrough
discoveries made during May-August 2005, is therefore promising and
timely. Here we report the results of such an investigation.

\section{Observations}

We have compiled all the SHBs localized to within error boxes
smaller than $10~ \rm arcmin^2$ with
$ \rm |Galactic~ latitude| > 20$, ending up with 5 SHBs:
790613, 000607, 001204, 021201 and 020531. We have defined the
following observational test to be carry out: we look for significant
luminosity overdensities in the fields of interest, in the $BVI$
bands, either in the form of a single luminous galaxy, or as an
overdensity of many fainter galaxies. We obtained spectra of brighter galaxies, 
determined their redshifts, and include this additional
information in our statistical analysis. A reexamination of the
afterglow search for SHB 020531 initially suggested the afterglow may
have been detected but overlooked (see \S \ref{sec 020531}). We have
therefore systematically explored the galaxy population only in the remaining four 
SHBs, as described below, and excluded SHB 020531
from the statistical analysis performed. We return to this point 
in \S \ref{sec 020531}.

\subsection{SHB 790613}

SHB 790613 was a short ($\sim48$ms) and hard GRB localized by the
IPN to within an extremely small error-box (in IPN standards; $0.7$
arcmin$^2$, \citealt{Barat84, Barat85}). Examination of optical
($BVI$) images, taken with the robotic $60''$ telescope (P60) at
Palomar Observatory, revealed a field dense with red galaxies both
within and outside of the error box. Four reddish galaxies all with
$i<20.5$ were within, or on the edge of, this small error box. This
is an apparently high density of galaxies even when compared to the
surrounding dense area. The imaging data show that these galaxies
have similar colors (Fig. 1) and suggest that they are probably
physically associated. In order to test whether this galaxy density
is unique we extract from the SDSS \citep{abazajian05} a catalog of
galaxies that cover $\approx 24~ \rm deg^2$ from regions with
Galactic extinction comparable to that in the direction of SHB
790613. We find that the probability to find four or more galaxies
with $m<20.5$, in a 0.7 arcmin$^2$ area, is $\approx 1\%$.

A query of the
NED database\footnote{\url{http://nedwww.ipac.caltech.edu/}}
has shown that this burst occurred within $6.5'$ from the cataloged
center of a rich Abell galaxy cluster (Abell 1892; richness class
$\Re=1$ [\citep{Abell58}]; cataloged redshift $z=0.09$ based on a
single galaxy redshift [\citep{Struble99}], and therefore somewhat
uncertain).
The Abell galaxy cluster catalog \citep{Abell58} contains 2712
galaxy clusters, of which 1894 ($\approx 70\%$) are considered to be
rich ($\Re \geq1$). The completeness of the catalog and the
resulting sky density of clusters is a strong function of Galactic
latitude. \citet{Abell58} reports $\approx0.08$ clusters per
degree$^2$ at Galactic latitude $40^{\circ}$ (the galactic latitude
of SHB 790613 is $37.6^{\circ}$), of which $\approx0.06$ are expected to be rich.
The chance probability to find a rich cluster within $6.5'$
($\approx0.1^{\circ}$) from a random point in this latitude is
therefore $P\approx \pi 0.1^2 \times 0.06 \approx 2 \times 10^{-3}$. We
conclude that this association is unlikely to be spurious,
even considering that we have checked for a similar association
(with negative results) in 4 other cases. Therefore we hereafter assume that
GRB 790613 occurred at $z=0.09$ and resided in one of the cluster early
type galaxies. At this redshift, the total isotrpic-equivalent energy 
emitted by this SHB is $6 \times 10^{49}$ erg (between $0.15-5$ MeV). The
burst luminosity is $\approx 1 \times 10^{51}$ erg sec$^{-1}$ cm$^{-2}$.

\begin{figure*}
\includegraphics[width=17cm]{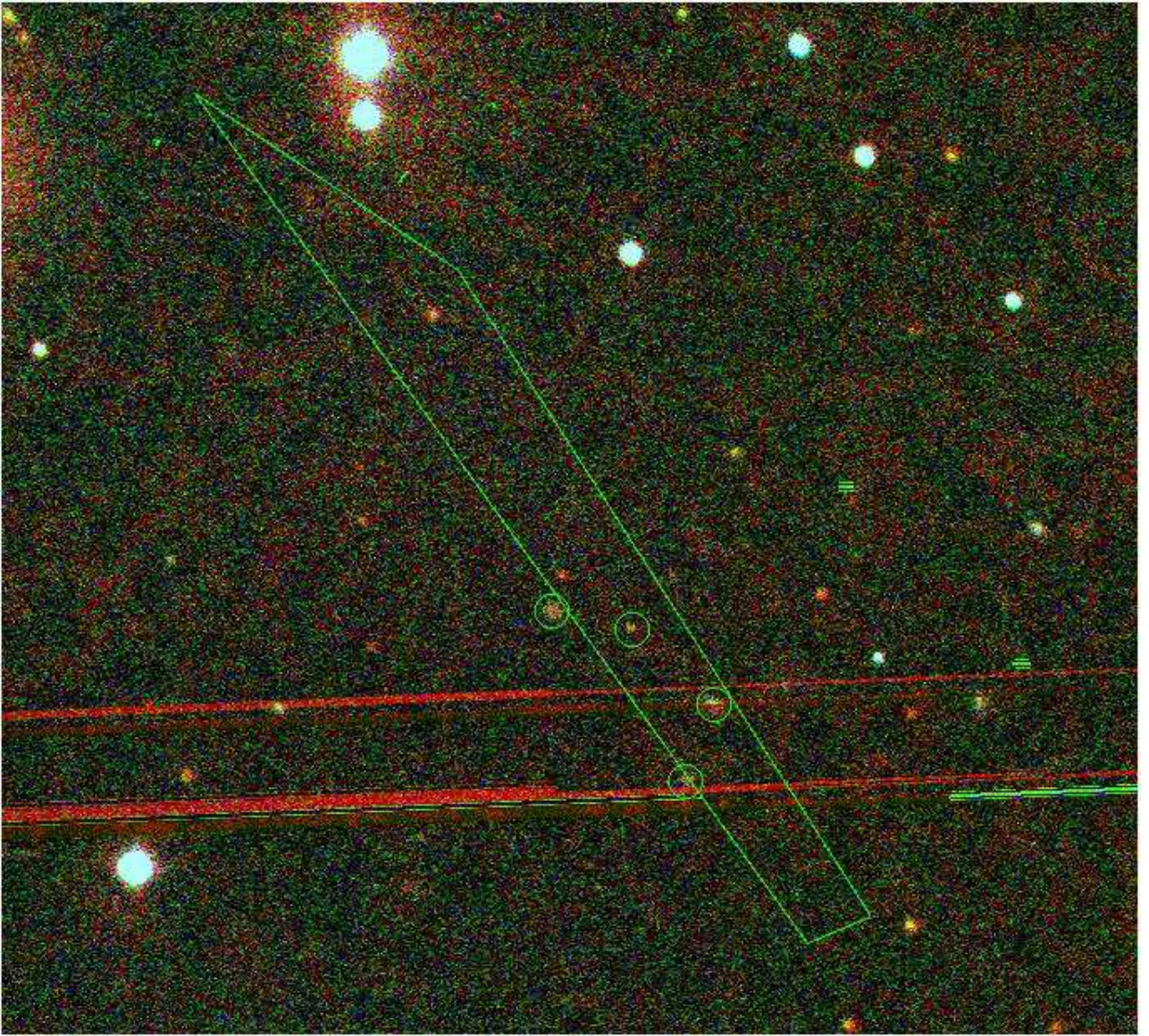}
\caption{A color image of the IPN error box of SHB 790613 (green polygon)
made from P60 $BVI$ imaging.
Four galaxies with apparently similar colors are visible
(green circles).}
\end{figure*}

Note that if the progenitor of this SHB was a NS-NS merger
\citep{Eichler89, Narayan92} then its host galaxy can also be
located outside of the error-box (1 arcmin corresponds to $\approx
100$ kpc at $z=0.09$, which a NS binary with a modest initial kick
velocity can travel before it merges).

\begin{deluxetable}{lllll}
\tablecaption{Journal of Observations}
\tablewidth{17cm}
\tablehead{
\colhead{SHB} & \colhead{Telescope} & \colhead{Instrument} & Exposure & UT Date\\
}
\startdata

Photometry: & & & &\\

\tableline

790613 & P60 & CCD & $B$ (450s), $V$ (450s), $I$ (450s) & Aug. 12, 2005\\
000607 & P60 & CCD & $R$ (720s) & Sep. 28, 2005\\
001204 & P60 & CCD & $B$ (450s), $V$ (450s), $I$ (450s) & Aug. 11, 2005\\
       & P60 & CCD & $V$ (1500s), $R$ (1500s), $I$ (900s) & Sep. 28, 2005\\
       & LCO100 & Tek5 & $V$ (180s), $I$ (90s) & Sep. 2, 2005\\
021201 & P60 & CCD & $B$ (600s), $V$ (600s), $I$ (600s) & Sep. 17, 2005\\

\tableline
\tableline

Spectroscopy: & & & &\\

000607-G1 & P200 & DBSP & Red+Blue (900s) & Feb. 6, 2005\\
001204-G1 & P200 & DBSP & Red+Blue (2700s) & Aug. 13, 2005\\
001204-G2 & P200 & DBSP & Red+Blue (10800s) & Sep. 9, 2005\\

\tableline

\enddata
\tablecomments{P60 = Palomar $60''$ robotic telescope. P200 = Palomar $200''$ Hale telescope.
LCO100 = Las Campanas Observatory $100''$ Du Pont telescope. Spectroscopic observations were
obtained using the double-beam spectrograph with the 600 line grating on the blue
side and 158 line grating on the red side, yielding a resolution of $\approx1$~\AA~ and
$\approx5$~\AA~, respectively.}
\end{deluxetable}

\subsection{SHB 000607}

SHB 000607 was well-localized by the IPN (error box area $5.6$
arcmin$^2$; \citealt{Hurley02}). In our previous study of this event
\citep{Nakar05} we have identified a bright galaxy (000607-G1). This
galaxy is the brightest galaxy in all the error boxes examined here.
We also determined its redshift to be $z=0.1405$ based on P200
spectroscopy (Fig. 2). In order to examine the probability to find
such a galaxy within the error box we first calibrated
its $R$ magnitude using P60 observations of \citet{landolt92} standard
stars, obtained during a photometric night. We then converted it to Sloan
Digital Sky Survey (SDSS)
$r$ and $i$ magnitudes using synthetic photometry applied to
the P200 spectrum, as described in \citet{Poznanski02}.
We obtain $r = 17.9\pm 0.1$ and $i=17.3 \pm 0.1$.

We estimate the expected local density of galaxies using the
Schechter function \citep{Schechter76} fit to the SDSS luminosity
function at $z\approx 0.1$ in the $r$ and $i$ bands from \citet{Blanton03}:
\begin{equation}\label{Eq LF}
\phi(L>L_0)=\int_{L_0}^\infty \phi^*(L/L^*)^\alpha e^{-L/L^*}dL/L^*
\end{equation}
where $\phi^*_r=5 \times 10^{-3} Mpc^{-3}$, $M^*_r=-21.2$,
$\alpha_r=-1.05$ and $\phi^*_i=5 \times 10^{-3} Mpc^{-3}$,
$M^*_i=-21.6$, $\alpha_i=-1$ (we assume the
standard cosmological model: $\Omega_\Lambda=0.7, \Omega_m=0.3$ and
$h=0.7$).

At $z=0.14$ (proper distance of $580$ Mpc) $L_{000607-G1}\approx
{1.3 \pm 0.1}~L_*$ (using the P200 spectrum for k-corrections) both
in $r$ and $i$. Thus  $\phi(L>L_{000607-G1}) \approx 6 \times
10^{-4} Mpc^{-3}$ and within a distance of $580$ Mpc there are about
$5 \times 10^5$ galaxies that are as bright or brighter than
$000607-G1$, implying an angular density of $3 \times 10^{-3}~ \rm
arcmin^{-2}$. Therefore the probability to find such a bright galaxy
within the $5.6~ \rm arcmin^2$ error box of GRB 000607 is
$\approx2$\% while the probability to find it anywhere within the
four error boxes searched (a total angular area of $21~ \rm
arcmin^2$) is $\approx7$\%. We conclude that the association between
this galaxy and GRB 000607 is significant ($\approx2\sigma$) though
not definitive. Given the pitfalls of a posteriori statistics, we do
not attempt to increase the statistical significance of the
association by further examination of this galaxy in search for
peculiar or noteworthy properties (colors, morphology, etc.).
At a redshift of $z=0.14$, the total isotropic-equivalent energy 
emitted by this SHB would be $2 \times 10^{49}$ erg (between $50-300$ keV) and
$2 \times 10^{50}$ erg between $0.15-5$ MeV. The corresponding
burst luminosity is $\approx 5 \times 10^{51}$ erg sec$^{-1}$ cm$^{-2}$.

The spectrum shape and emission line strength of $000607-G1$
indicate an intermediate spiral galaxy, perhaps of type Sb. Using
the \citet{Kennicutt98} relation we estimate a star formation rate
of $\approx0.3$ M$_{\odot}$ y$^{-1}$ from the observed H$\alpha$
luminosity, similar to that of the host galaxy of SHB 050709
\citep{Fox05}.

\begin{figure*}
\includegraphics[width=17cm]{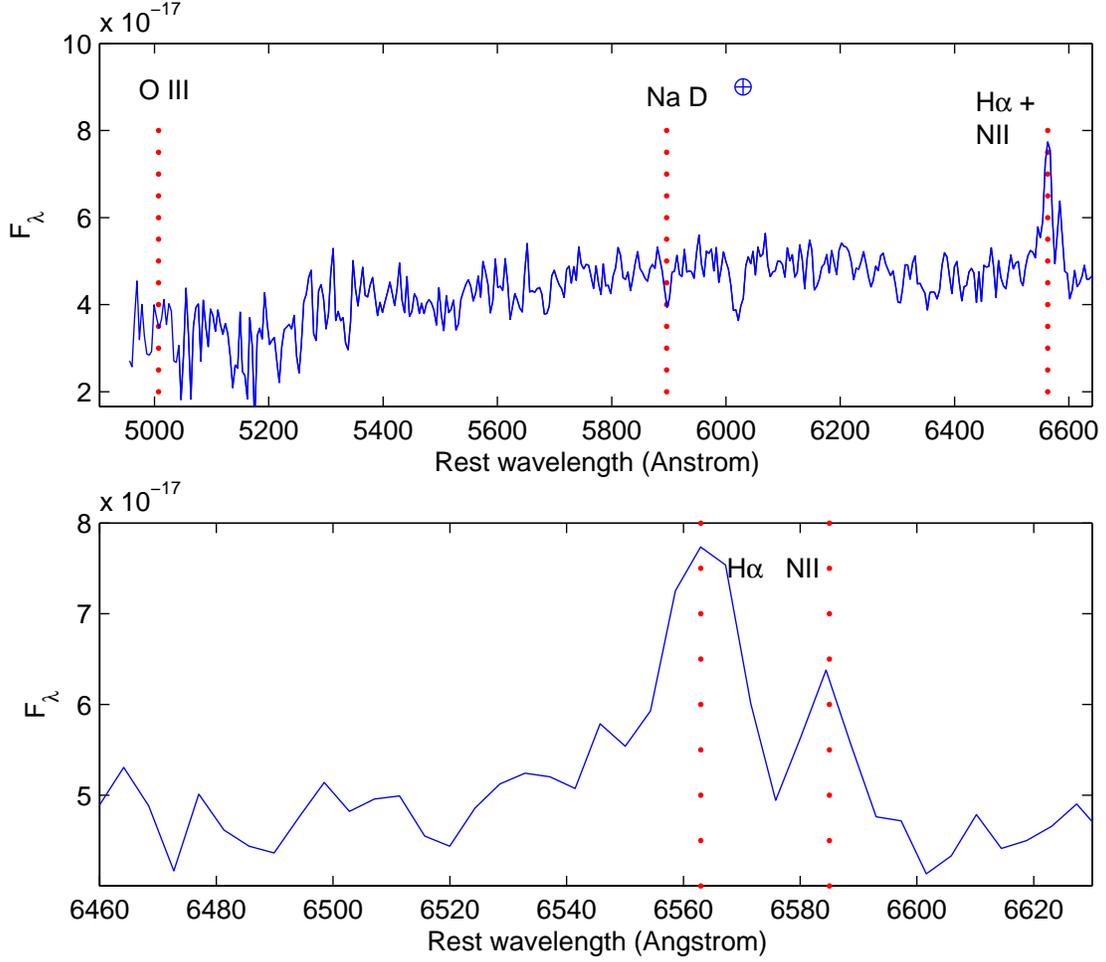}
\caption{The spectrum of the bright galaxy 000607-G1, identified as
the likely host of SHB 000607. We determine the redshift from
H$\alpha$ ($6563$~\AA) and NII ($6585$~\AA) in emission (see lower
panel for a detailed view) and Na D ($5896$~\AA) in absorption. OIII
($5007$~\AA) is weak or absent. Telluric absorption is present near
$6000$~\AA~ (circled cross). The spectrum shape and emission line
strength indicate an early or intermediate spiral galaxy, perhaps of
type Sb. Using the \citet{Kennicutt98} relation we estimate a star
formation rate of $\approx0.3$ M$_{\odot}$ y$^{-1}$ from the
observed H$\alpha$ luminosity. This is actually an upper limit as
the H$\alpha$ line is contaminated by nearby NII ($6550$~\AA)
emission (bottom panel). The strength of the resolved NII line at
$6585$~\AA~ suggests this contamination is small ($<20\%$), but the
low S/N and low resolution of the P200 spectrum preclude a more
accurate determination.} \label{fig:000607-spec}
\end{figure*}

\subsection{Additional Bursts}

\subsubsection{SHB 001204}

SHB 001204 was localized by the IPN  to within a $6$ arcmin$^2$
error box (\citealt{Hurley02}). In our previous study of this event
\citep{Nakar05} we have identified the brightest blue galaxy in this
field, $001204-G1$. We have since acquired P200 spectroscopy
indicating that this galaxy is at $z=0.31$ (Fig. 3). Finding a
galaxy of this luminosity at this redshift in a random sky patch
with an area similar to the SHB error box is not unexpected. P60 BVI
imaging, as well as VI photometry obtained at the $100''$ Du Pont
telescope at Las Campanas Observatory, reveals 5 more red galaxies
within the error box (Fig. 4), brighter than $r \approx 21$. P200
spectroscopy of the brightest of these ($001204-G2$) indicates it is
at $z=0.388$ (Fig. 5). In order to test whether this galaxy density
is unique we extract from the SDSS \citep{abazajian05} a catalog of
galaxies that cover $\approx 15~ \rm deg^2$ from regions with
Galactic extinction comparable to that in the direction of SHB
001204. We find that the average number of galaxies with $r<21$, in
a 6 arcmin$^2$ area, is $\approx6$. We therefore conclude that the
error box of SHB 001204 does not contain any uncommonly bright
galaxy or a galaxy overdensity down to this limit.

\begin{figure*}
\includegraphics[width=17cm]{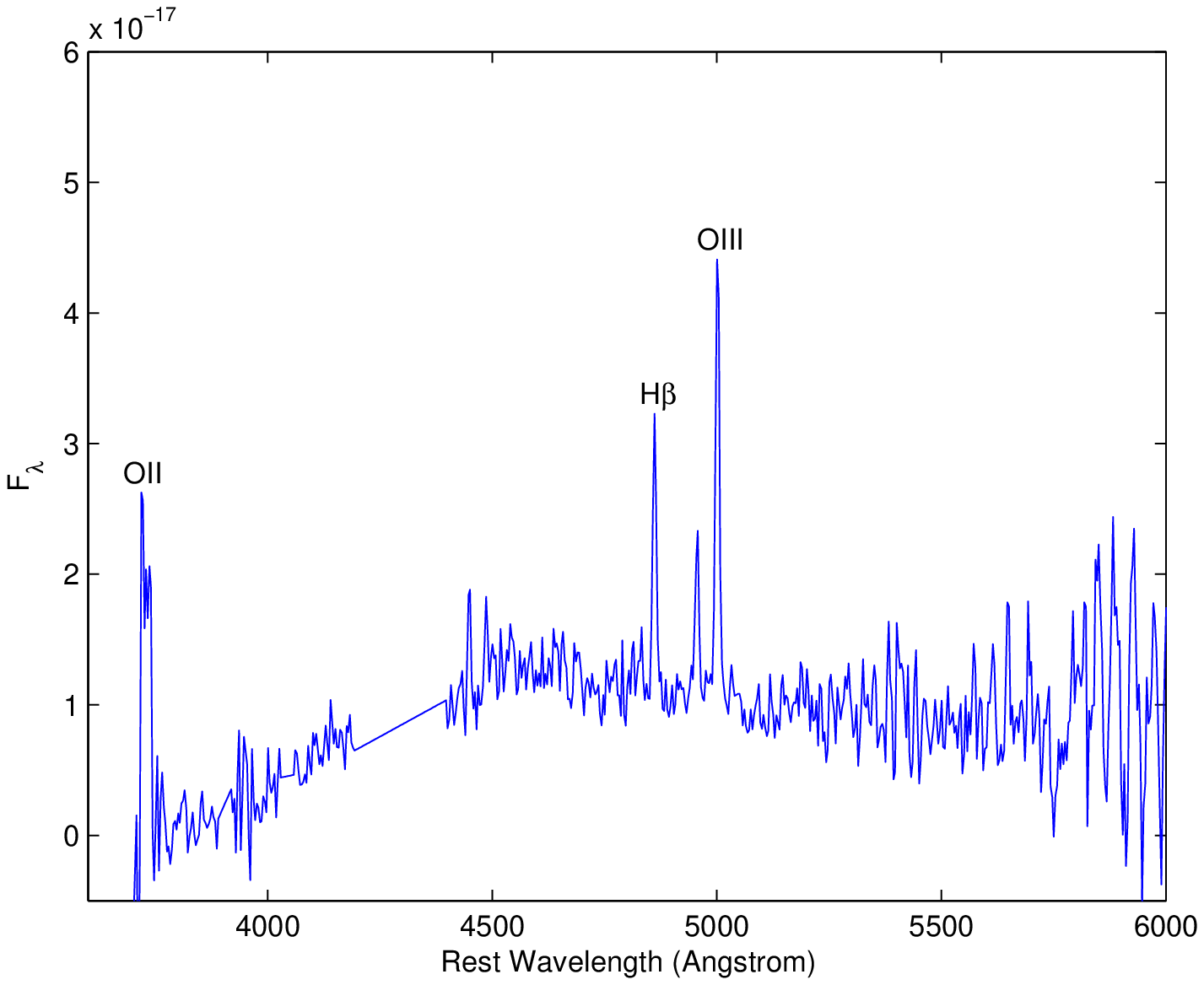}
\caption{P200 spectrum of $001204-G1$. Areas affected
by strong sky line residual have been excised.
Prominent emission lines
(OIII 5007+4949~\AA, H$\beta$, OII 3727~\AA) indicate
$z=0.31$. The emission line strength and overall shape
are consistent with those of a late spiral (Sbc or similar).}
\end{figure*}

\begin{figure*}
\includegraphics[width=17cm]{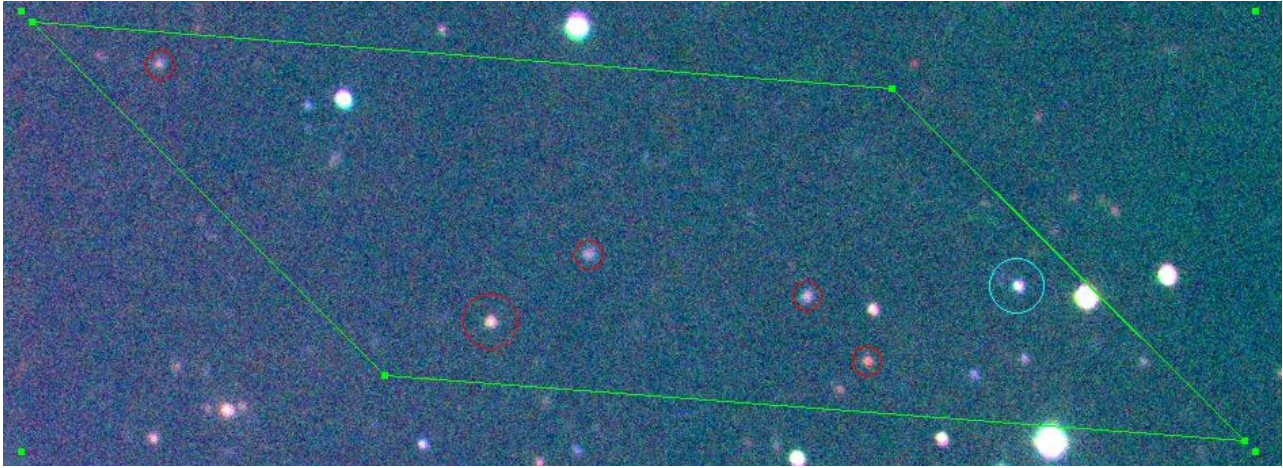}
\caption{A color image of the IPN error box of SHB 001204 (green)
made from LCO $VI$ and P60 $R$-band photometry.
The blue galaxy discussed by Nakar et
al. (2005; $001204-G1$; cyan circle) is at $z=0.31$, while the
brightest red galaxy in the field ($001204-G2$; large red circle) is
at $z=0.388$ (Fig. 6). Additional red galaxies
are indicated by smaller red circles.}
\label{fig:000607-spec}
\end{figure*}

\begin{figure*}
\includegraphics[width=17cm]{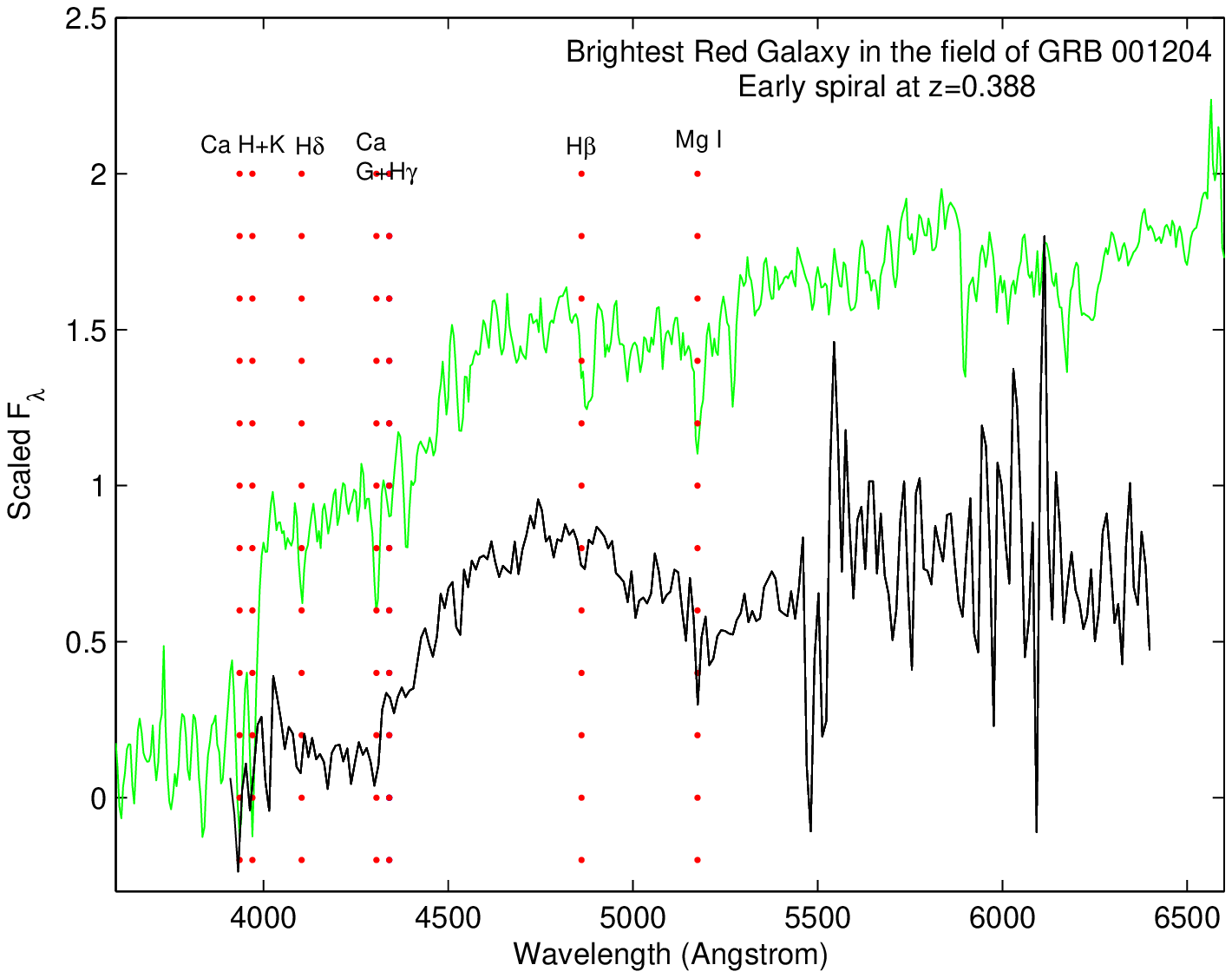}
\caption{P200 spectrum of $001204-G2$ (black). We identify
absorption lines of Ca (G, H and K bands) H (H$\delta$, H$\gamma$
and H$\beta$) and Mg I, indicating $z=0.388$. Comparison with an Sa
template spectrum from \citealt{Kinney96} (green) suggests this is
an early spiral galaxy.}
\end{figure*}

Valuable information can be extracted also from a null detection.
Bases on the association of SHBs with an older stellar population
(at least a few time $10^8$ years old) we can assume that the rate
of SHBs follows either the blue or red luminosity, most of which is
located within relatively bright galaxies. This is not the case for
younger stellar populations, traced by the UV light, a larger
fraction of which is associated with intrinsically less luminous
galaxies. As discussed in \citet{Nakar05} this assumption implies
that likely host galaxies should be more luminous than
$\approx 0.33[0.02]$L$_{*}$ at $1[2]\sigma$ in these colors. The
brightest galaxy for which we did not obtain a redshift in the error
box of GRB 001204 has $r \approx 20.3$ corresponding to a minimal
redshift of $z=0.25[0.06]$ at $1[2]\sigma$, implying a lower limit
on the isotropic bolometric energy release in $\gamma$ rays of
$E_{iso}=5[0.25] \times 10^{50}$ erg.

\subsubsection{SHB 021201}

SHB 021201 has the largest error box in our sample, it was localized
by the IPN  to within $9$ arcmin$^2$ \citep{hurley02c}. In our
previous study of this event \citep{Nakar05} we have identified what
appeared to be a bright blue galaxy in a Palomar Observatory Sky
Survey 2 (POSS2) $B$ plate of this field obtained on 1995 (Fig. 6).
However, P200 spectroscopy of this source that we have since
obtained, indicates that it is an M star, and indeed, it appears to
be point like in $R$ and $I$ POSS2 plates obtained during 1997-1998,
as it does on older plates from the USNO plate archive
\footnote{\url{http://www.nofs.navy.mil/data/fchpix/}}. The extended
appearance of this source is therefore either related to a plate
defect, or suggests that this star has undergone an unusual
ejection/illumination event around 1995.

\begin{figure*}
\includegraphics[width=17cm]{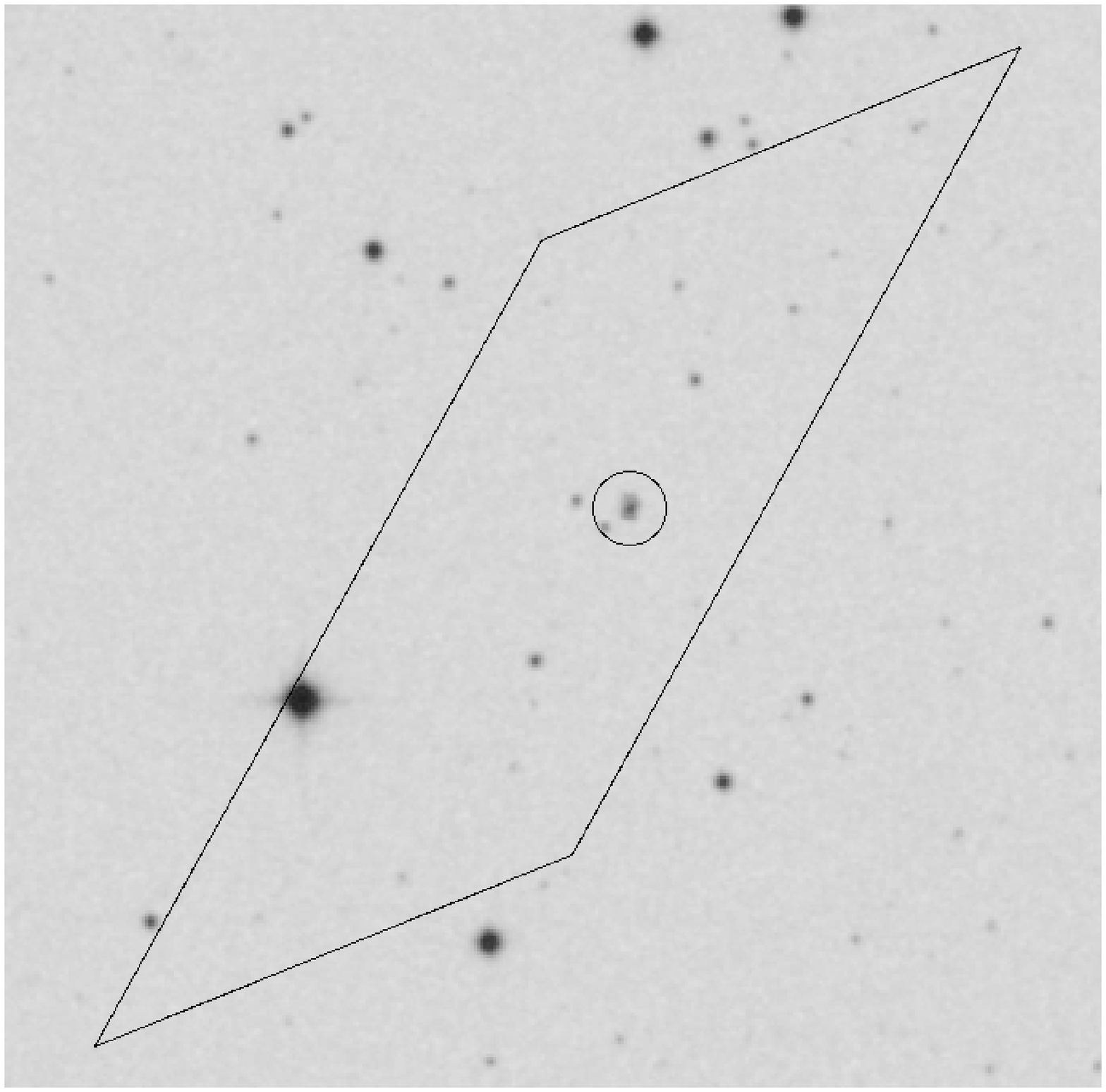}
\caption{Reproduction of a section of the POSS2
B plate extracted from the ESO digital sky survey
archive (www.eso.org/dss) showing the location
around the error box of SHB 021201 (black polygon).
An apparently resolved object (circled), previously assumed
to be a galaxy, is actually an M star.}
\end{figure*}

There are no other bright galaxies in this error box, which is
included in the Sloan Digital Sky Survey Data Release 4 (SDSS DR4;
released in July 2005, and not available during our previous analysis).
The galaxy content of this error box is sparse even when compared
to that of SHB 001204, and is not consistent with a local overdensity
of galaxies, down to the SDSS limit.
The brightest galaxy included in the SDSS database has $r=20.25$
resulting in a lower limit similar to that obtained above for SHB 001204.
Since both
events have a similar energy output \citep{Nakar05} the lower limit on
the energy release in $\gamma$ rays for SHB 021201 is
$E_{iso}=5[0.25] \times 10^{50}$ erg (at $1[2]\sigma$) as well.

\subsubsection{SHB 020531}\label{sec 020531}

This event was detected by {\it HETE-2} \citep{Ricker02}, and its
localization was improved several times by analysis of {\it HETE-2}
data in conjunction with other spacecrafts from the IPN (final
position given by \citealt{Hurley02b}). Sensitive X-ray observations
with the {\it Chandra} observatory were undertaken by
\citet{Butler02a} and revealed numerous X-ray sources within the
initial IPN error box, one of which (source $\#48$,
\citealt{Butler02b}) showed significant fading, and was considered
to be a viable candidate for the X-ray afterglow of this event
\citep{fkw02}. However, the final refinement of the IPN localization
no longer included source $\#48$. We noted that the
strongest {\it Chandra} source ($\#0$, \citealt{Butler02a}),
hitherto outside of previous IPN localizations, was included in the
latest revised box. This source shows
significant decay and is coincident with an optical
source detected on DSS plates \citep{Butler02a}. 
Sensitive radio observations taken as part of the afterglow search
\citep{frail02} place a $5\sigma$ upper limit of $250\mu~$Jy on
radio emission at $5$ or $8$ Ghz from this location, indicating this
source is not a radio-loud AGN.

In view of the tendency of well-localized SHBs to reside in
apparently bright hosts \citep{Kulkarni05,Berger05,Bloom05} and the
detection of strong and long-lasting X-ray afterglows from SHBs
\citep{Fox05,Berger05,Romano05} we initially thought that
{\it Chandra} source $\#0$ may be the X-ray afterglow of SHB
020531, and that the underlying optical source may be its host.
We therefore did not pursue a ``blind'' luminosity overdensity
test for this field as described above. 
However, it turns out that optical spectra of this source show it is
an AGN (Butler et al. private communication) naturally explaining its
X-ray variability, and discrediting its association with SHB 020531. 
Since this burst was not, initially, part of our plan for statistical study of 
IPN SHBs, its late inclusion in our sample will hinder our attempt 
to minimize effects of a posteriory statistics. We therefore exclude
it from the statistical sample discussed here. 

\section{Discussion}

The observations and analysis reported in the previous section allow
us to significantly increase the number of SHBs for which redshift
and host galaxy information is now available (Table 2). We now
turn to investigate what can be learned from this extended sample of
events.

\subsection{Host galaxies}

We have compiled the properties of known SHB host galaxies
in Table 2 and Fig. 7 (dark blue histogram).
We assign E/S0 hosts for cluster events (790613 and 050813).
Inspection of the distribution of observed Hubble types indicates
a large fraction of early type hosts, with some
events located within later hosts. The apparent ubiquity
of SHBs in galaxies of many types calls to mind another
type of explosive phenomenon, namely, supernovae (SNe)
of type Ia, as already mentioned by \citet{Fox05},
\citet{Berger05}, and \citet{prochaska05}.

\begin{deluxetable}{lllll}
\tablecaption{Host galaxies and redshifts of SHBs}
\tablewidth{17cm}
\tablehead{
\colhead{SHB} & \colhead{Redshift} & \colhead{Host Galaxy} & Association  & Reference\\
              & $z$                & Type                  & significance &
}
\startdata

790613 & 0.09 & E/S0 & $\sim 3\sigma$ & This work \\
000607 & 0.14 & Sb   & $\sim 2\sigma$ & This work \\
050509b & 0.22 & E/S0 & $3-4\sigma$ & \citealt{Bloom05,Kulkarni05}\\
       &      &      &  & \citealt{Castro05a,Gehrels05} \\
050709 & 0.16 & Sb/c &  Secure & \citealt{Fox05}\\
050724 & 0.26 & E/S0 & Secure & \citealt{Berger05,prochaska05} \\
050813 & 0.72 & E/S0 & - & \citealt{Gladders05,berger05b} \\
\tableline
001204 & $>0.25[0.06]$ & - & $1[2]\sigma$ & This work \\
000607 & $>0.25[0.06]$ & - & $1[2]\sigma$ & This work \\
\enddata
\end{deluxetable}

\begin{figure*}
\includegraphics[width=17cm]{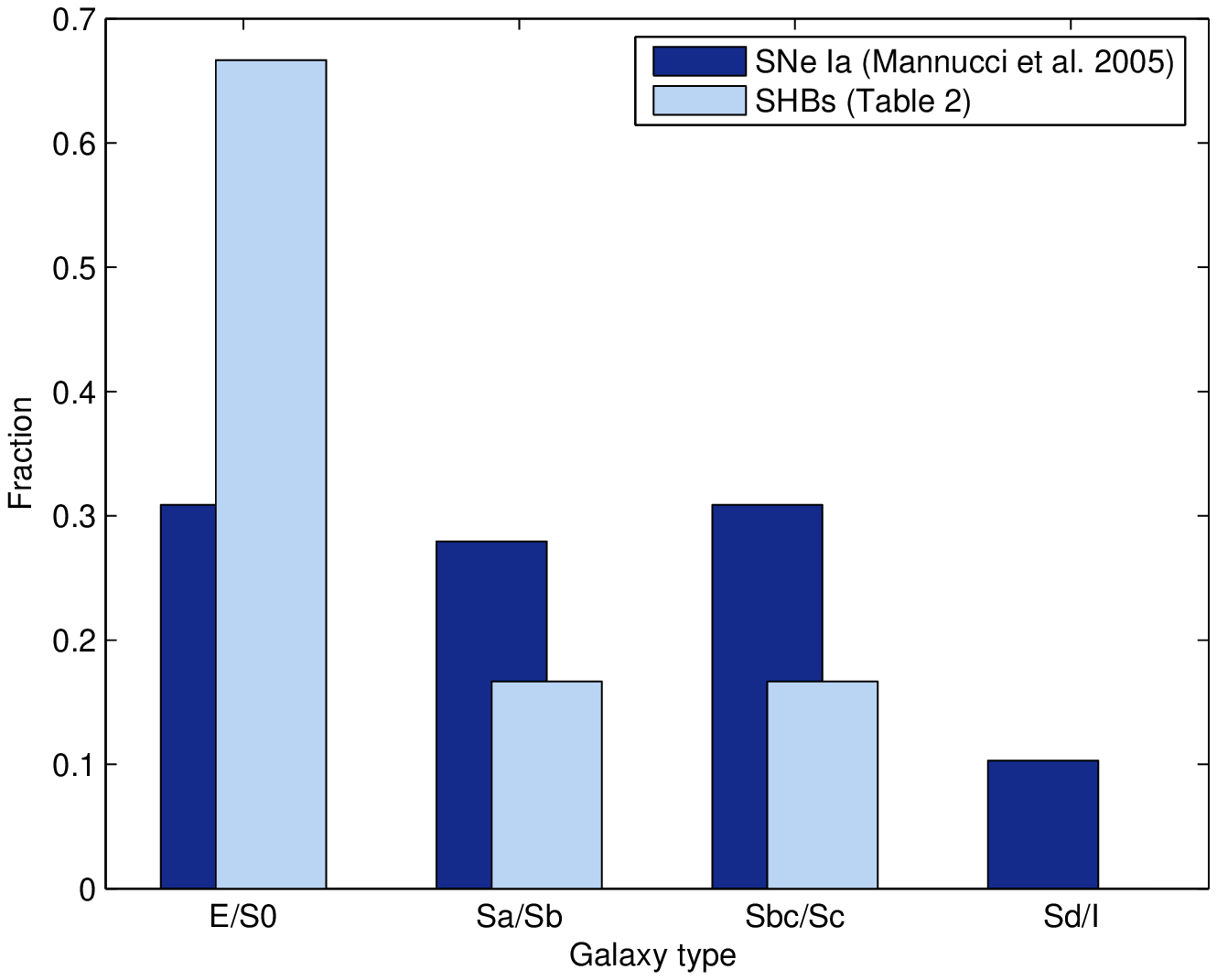}
\caption{A comparison between the host galaxy types of 
SHBs (from Table 2) and SNe Ia
\citep{mannucci05}. The fraction of SHBs in early type galaxies is
significantly larger than the fraction of SNe Ia observed in such
galaxies in the nearby Universe, indicating that the progenitor
systems of SHBs are probably longer-lived than those of SNe Ia. }
\end{figure*}

SNe Ia are believed to result from a thermonuclear runaway explosion
of a white dwarf star, at or near the Chandrasekhar mass, triggered
by accretion from, or merging with, a binary companion. These SNe
occur in galaxies of all types, including early type galaxies with
little or no recent star formation, and in this respect they appear
to resemble SHBs. However, comparing the observed Hubble type 
distribution of host galaxies of SHBs (Table 2) and SNe Ia (from
\citealt{mannucci05}; light grey histogram in Fig. 7), clearly
suggests a difference between these two phenomena. While SNe Ia
indeed occur in E/S0 galaxies, the majority of events explode in
spirals, and more than half in galaxies later than Sa. Most SHBs, on
the contrary, appear to occur in early type hosts. The probability
that the observed SHB host galaxies are drawn from the SN Ia host
distribution of \citet{mannucci05} is small ($P\sim7\%$).
\citet{mannucci05} calculate also the rate of SNe Ia per
unit $K$-band luminosity (an indicator of the old stellar mass).
They find that this normalized rate increases by more than an order of
magnitude from E/S0 to Irr galaxies, implying that a significant
fraction of the progenitors of SNe Ia are associated with young
stars. This is not the case for SHBs. In fact, the current census of
SHBs is consistent with all the progenitors coming from an older
stellar population, abundant in E galaxies and bulge/spheroid
components of early and intermediate spirals, and accounting for
$\sim75\%$ of the total current stellar mass \citep{fukugita98}.
Thus, the comparison between the observed distributions of SHBs and
SNe Ia host galaxies (Fig. 7) indicates that SHBs originate from
systems that have longer typical lifetimes than those of SNe Ia.

The lifetime of SN Ia progenitor systems (often parameterized by the
typical delay time $\tau_{Ia}$ between star formation and SN
explosion) is currently an open observational question. If all SNe
Ia originate from the same class of progenitors one can assume a
unimodal delay-time function parameterized by a typical delay time
and a distribution around this mean. Several authors used the
observed redshift evolution of SN rates and their redshift
distributions, along with a prescription for the star formation
history of the Universe \citep{tonry03, barris05, strolger04}, and
the host properties of SNe Ia \citep{mannucci05} to constrain the
typical delay time. \citet{tonry03}, \citet{barris05} and
\citet{mannucci05} indicate relatively short delay times
($\tau_{Ia}\sim1$ Gyr) while the analysis of \citet{strolger04}
prefers a much longer delay time ($\tau_{Ia}>2$ Gyr). The combined
analysis of SNe Ia in both field and cluster environments
\citep{galyam04,maoz04} indicates that short delay times
($\tau_{Ia}\simlt 1$ Gyr) are in conflict with popular star
formation history models, while long delay times ($\tau_{Ia}>3$ Gyr)
are inconsistent with SNe Ia being the source of metals in the
intra-cluster medium (ICM) of rich galaxy clusters. The current
consensus may lean toward a shorter delay time. The long delay time
advocated by \citet{strolger04} appears to be less convincing
(critically depending on estimates of the faint SN discovery
efficiency in deep {\it HST} data [see \citealt{barris05}]) and
cannot be reconciled with the straightforward observations of
\citet{mannucci05} in nearby galaxies.

In any case, since a larger fraction of SHBs occur in early type
galaxies compared with SNe Ia, they must have a significantly longer
delay time, on average, than the typical delay time of SNe Ia, i.e.,
of order several Gyr, even if we adopt a shorter delay time for SNe
Ia ($\tau_{Ia}\approx1$ Gyr). This finding disfavors the binary NS merger
model for SHBs if NS mergers are indeed dominated by systems with very short 
($\approx$~My) typical delay times (e.g., 
\citealt{tutukov93, Belczynski01, perna02}).

\citet{mannucci05} and \citet{scannapieco05} consider a
two-component model for SNe Ia. A long-lived component which comprises
the entire SN Ia population in early type galaxies,
and a ``prompt'', short-lived component, proportional to the star formation 
rate, dominant in late-type galaxies. This model naturally explains
the results of \citet{mannucci05} and skews the combined rate
toward shorter delay times.  Interestingly, the distribution of SHB
host types is consistent with the expected distribution of the
long-lived SN Ia component, which accounts for the majority of
events in E galaxies, about $50\%$ in S0/a/b galaxies, about $20\%$
in Sbc/d galaxies, and hardly contributes to the rate is the latest
Ir galaxies \citep{mannucci05}. Such a component would originate
solely from the oldest stellar population, and would have a typical
age of order $10$ Gyr.

It must be noted that our analysis assumes that our search and analysis
procedures are not strongly biased toward discovering early-type
hosts. This is a fair assumption for single luminous galaxies, as
demonstrated by the fact that the luminous galaxy we uncovered
(000607-G1) is an intermediate spiral and not a red, early-type
galaxy. A bias may exist for fainter galaxies, since red galaxies
are more clustered than blue ones, and a faint red galaxy that would
have escaped notice as a single galaxy, may be detected by us due to
its association with a galaxy overdensity. attempts to quantify and
correct for this bias will have to await the assembly of larger
future samples of SHB hosts.

\subsection{Redshift distribution}

An alternative approach to constrain the lifetime distribution of SHB progenitors,
$P(\tau_{SHB})$, is to use the observed SHB redshift
distribution in conjunction with the observed flux distribution \citep{ando04,Guetta05}.
These observed quantities are determined by three components:
the {\it intrinsic} redshift
distribution, the intrinsic luminosity function and observational
selection effects (detection threshold and efficiency, ability to determine the
redshift and so on). So, using two observed distributions we attempt to
solve for two unknown intrinsic functions (assuming that the observational selection
effects can be accounted for) of which only one is expected to be related
directly to the progenitor lifetime (the {\it intrinsic} redshift
distribution). It therefore might be expected that this approach will
prove less constraining than the more direct
investigation based on host galaxy type. On the other hand, it takes advantage of
the large amount of data collected by the BATSE experiment, and is thus
worth exploring.

\citet{Guetta05} predict the {\it observed} redshift
distribution of SHBs assuming that the luminosity function is a
broken power-law and that the {\it intrinsic} redshift distribution
is a convolution of the lifetime distribution with the star
formation rate history (a similar approach was used to probe
SN Ia delay times, e.g., by \citealt{Madau98} and \citealt{galyam04}). 
The observed flux
distribution and the detection threshold are taken from the BATSE
catalog. \citet{ando04} conducted a similar study but considered 
a wider range of luminosity functions, lifetime distributions, and 
star formation histories. 
In Fig 8 (right panel) we depict the redshift distribution
calculated by \citet{Guetta05} under the assumption 
$P(\tau_{SHB})\propto1/\tau_{SHB}$ deduced from 
the observed sample of Galactic NS-NS binaries \citep{Champion04}. 
Fig 8 also shows the observed distribution of
the four bursts detected by {\it Swift} (which is expected to have a
detection threshold comparable to BATSE) and {\it HETE-2}. We do not
include in this analysis the IPN bursts (Fig. 8, left panel) for
which the detection threshold is expected to be much higher.
Although the sample is very small, the null hypotheses that the
observed redshift distribution is drawn from the predicted one is clearly
($>3\sigma$) rejected. Three out of the four bursts are at $z<0.3$,
a region that includes only $7\%$ of the model distribution.
Including only the three {\it Swift} bursts reduces the rejection
significance to $\approx2\sigma$. Comparison with the range of models
presented by \citet{ando04} yields similar results. This suggests that the lifetime
distribution of SHBs favors {\it longer} lifetimes than those
inferred from the observed distribution of Galactic NS-NS binaries. Taken at
face value, the observed properties of well-localized SHBs are
difficult to reconcile with the latest predictions based on the
popular NS-NS merger progenitor model.

Strong observational selection effects against the detection of
NS-NS binaries with short merger timescales are expected to skew the
observed merger timescale distribution toward longer values.
Correcting for this bias would make the NS-NS binary model
predictions even more discrepant with the observed SHB redshift
distribution. A selection effect that might prevent the detection of
old binary systems may be introduced by the (relatively) short 
lifetimes of pulsars, which are used to 
detect such systems. It might be that there is a large number of old
($\sim 10$ Gy) NS binaries in our Galaxy composed of two ``dead''
pulsars. Nevertheless, the currently known NS binary sample contains too
many systems with a relatively short lifetime ($\lesssim 1$ Gyr)
that are disfavored by the observe redshift distribution of SHBs, as
well as by the galaxy type analysis of the previous section.

What possible explanations may allow to reconcile the observed
distributions with the predictions from NS-NS merger models? It may
be that the {\it Swift} threshold for SHB detection is
higher than that of BATSE (perhaps due to the hard SHB spectrum,
while {\it Swift} is more sensitive in softer bands).
The recent {\it Swift} SHB detection statisctics seems to indicate,
though, that the SHB detections efficacies of both instruments are
actually comparable.
Yet another resolution may be that the true SHB luminosity
function is very different from the range considered by \citet{ando04}
and \citet{Guetta05}, which fits the BATSE data well, or a combination 
of these two effects. 
Another possibility is that this is a result of the small 
sample available, making our results vulnerable to selection
effects (the significance levels reported throughtout 
properly account for the statistical impact of the small sample size). 
Finally, it is possible that the
true typical lifetime of NS binaries is much longer than currently
indicated by the very few known NS binary systems. These issues
certainly merit further investigation.

\begin{figure*}
\includegraphics[width=17cm]{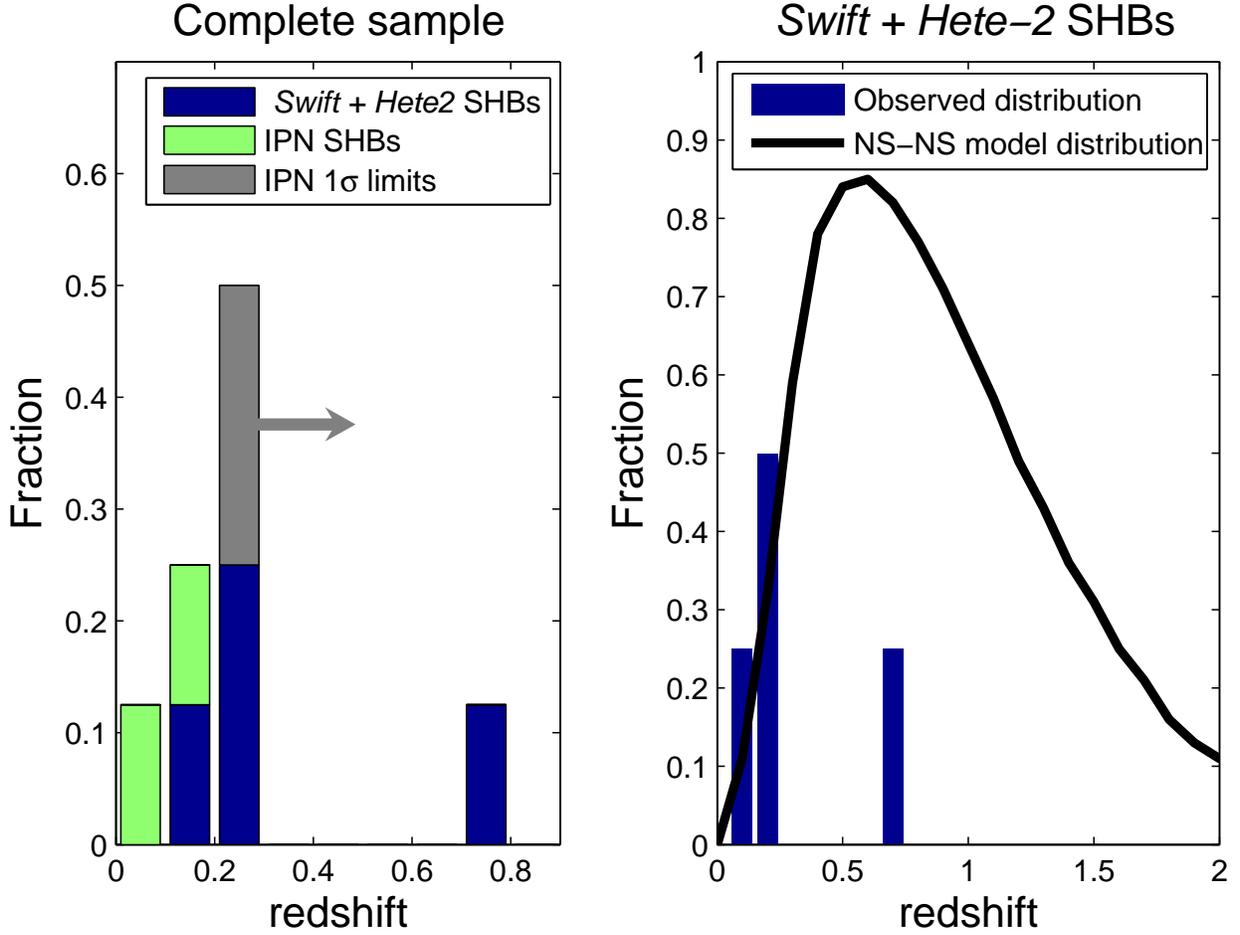}
\caption{The redshift distribution of SHBs. {\bf Left:} the observed
redshift distribution of SHBs from {\it Swift} + {\it HETE-2} (blue)
and the IPN (green). We also mark the $1\sigma$ lower limits on the
redshift of two additional IPN SHBs (grey). {\bf Right:} Comparison
between the {\it Swift} + {\it HETE-2} SHB sample, and the model
predictions (solid line) from \citet{Guetta05}, calculated assuming
the observed distribution of merger delay times for NS-NS binaries.
}
\end{figure*}

\section{conclusions}

We have analyzed new and archival observations of the fields of 
well-localized IPN SHBs. Using these data, we determine that
SHB 790613 is associated with a rich galaxy cluster, probably at
$z=0.09$, and that SHB 000607 has likely occurred in a $z=0.14$
luminous Sb galaxy. We use our null results
for the fields of two additional SHBs to set a lower limit on the
most likely redshift of these events.

We combine our new findings with published data for four recent SHBs
detected by {\it Swift} and {\it HETE-2}, and examine the properties
of this extended sample of events. Focussing on the distribution of
host galaxy types, as well as on the SHB redshift distribution. We
arrive at the following conclusions:

$\bullet~$ SHBs apparently occur in host galaxies of all types, as
do SNe Ia. However, SHBs appear to favor early-type hosts compared
to SNe Ia, a strong indication that they originate from a population
of progenitors which has a longer lifetime, on average. Even if we
adopt the shorter values derived for the typical SN Ia delay time
($\sim1$ Gyr), the progenitors of SHBs appear to require a longer delay
time, of order several Gyr. This finding disfavors the binary NS merger
model for SHBs if NS mergers are indeed dominated by systems with very short 
($\approx$~My) typical delay times (e.g., \citealt{tutukov93, Belczynski01, perna02}). 
The current sample of SHB
host galaxies is consistent with SHBs originating solely from the
older spheroid/bulge stellar population, suggesting a typical life
time of $\sim10$ Gyr in this case.

$\bullet~$ The observed redshift distribution of SHBs appears 
inconsistent with recent NS-NS merger model predictions based on BATSE
luminosity functions, universal star formation rate histories, and,
specifically, the distribution of delay times indicated by the known
sample of NS binaries in our galaxy \citep{ando04,Guetta05}. 
The binary NS-NS merger model is disfavored by the
data, as it predicts too many short-lived/high redshift progenitor
systems. Observational selection biases against binary NS-NS systems
with long merger time scales and/or against the detection of
late-type host galaxies, and/or modified 
SHB luminosity function, perhaps coupled with {\it Swift} being
less sensitive than BATSE, are required for the NS-NS model to remain viable.
Increased statistics combined with additional modeling work would
probably shed more light on this issue.

$\bullet~$ Long SHB progenitor lifetimes, if born out by further
investigations, imply that SHBs should hardly occur above $z\sim 1$.
Additionally, if the typical SHB delay time is several Gy, then we
predict that the sample of SHBs discovered at $0.5<z<1$ will show a 
strong preference to reside in the centers of galaxy clusters. At these
redshifts, the age of the Universe becomes comparable to the SHB lifetime,
so such events would occur where the oldest stellar population is
concentrated. According to hierarchical galaxy formation models, the first
stars, in the first galaxies, tend to form at the most 
extreme initial density peaks, which later evolve to be the locations of 
rich galaxy clusters. 

$\bullet~$ Our expanded sample of SHBs with known or probable redshifts implies
that a large fraction of SHBs occur at a low redshifts ($z<0.3$; within a
distance of $\sim 1$ Gpc). This is true even when we consider only the Swift
sample for which the threshold is similar to BATSE. This typical redshift is
smaller than previous estimates (e.g. \citealt{ando04,Guetta05}) resulting
in a higher {\it observed} local rate of $> 10 \rm ~Gpc^{-3} ~yr^{-1}$,
based on the observed BATSE rate of $\sim 170$ SHBs per year over the entire sky. 
This is a strict lower limit since it does not include dim bursts that were
missed by BATSE. It also does not account for possible beaming corrections which
might be significant \citet{Fox05}. If SHBs are NS-NS or NS-BH mergers
then this rate predicts a detection of the gravitational waves produced
during such mergers by advanced LIGO.     

\section*{Acknowledgments}

We thank S. Ando, N. Butler, A. Coil, B. Gerke, S. Phinney and C. Steidel for help and
advice. A.G. acknowledges support by NASA through Hubble Fellowship
grant \#HST-HF-01158.01-A awarded by STScI, which is operated by
AURA, Inc., for NASA, under contract NAS 5-26555. E.N. was supported
by a senior research fellowship from the Sherman Fairchild
Foundation. E.B. acknowledges support by NASA through Hubble
Fellowship grant \#HST-HF-01171.01-A awarded by STScI, which is
operated by AURA, Inc., for NASA, under contract NAS 5-26555.
AMS is supported by the NASA Graduate Student Research Program. SRK's
research is supported by NSF and NASA. This research has made use of
the NASA/IPAC Extragalactic Database (NED)  which is operated by the
Jet Propulsion Laboratory, California Institute of Technology, under
contract with the National Aeronautics and Space Administration.
This research has made use of the USNOFS Image and Catalogue Archive
operated by the United States Naval Observatory, Flagstaff Station
(http://www.nofs.navy.mil/data/fchpix/).

\bibliographystyle{apj}
\bibliography{ms}


\end{document}